\documentclass[11pt]{article}
\usepackage[utf8]{inputenc}
\usepackage[T1]{fontenc}
\usepackage{amsmath,amssymb}
\usepackage{mathptmx}
\usepackage{graphicx}
\usepackage[table]{xcolor}
\usepackage[letterpaper,margin=1in]{geometry}
\usepackage{enumitem}
\usepackage{booktabs,tabularx}
\usepackage{microtype}
\usepackage[font=small,labelfont=bf]{caption}
\usepackage{titlesec}
\usepackage{fancyhdr}
\usepackage{placeins}

\usepackage[hidelinks]{hyperref}
\titleformat{\section}{\centering\bfseries\scshape}{\thesection.}{0.6em}{}
\titleformat{\subsection}{\bfseries}{\thesubsection}{0.6em}{}
\titlespacing*{\section}{0pt}{1.6em}{0.8em}
\titlespacing*{\subsection}{0pt}{1em}{0.4em}

\begin{document}
\thispagestyle{plain}
\begin{center}
{\LARGE Cursive: The Trace from the Curse of Dimensionality}\\[1.3em]
{\normalsize\scshape By Hao Chen}\\[0.5em]
{\normalsize\itshape Department of Statistics, University of California, Davis}\\[0.4em]
{\footnotesize\texttt{hxchen@ucdavis.edu}}
\end{center}
\vspace{0.8em}
\begin{abstract}
Modern data are increasingly high-dimensional or non-Euclidean. As dimension grows, new statistical patterns can emerge in the relations among observations, while a conventional statistical summary may fail to retain the signal they carry. This paper names and organizes a research program around this observation, calling it \emph{Cursive}. Cursive asks which relational information is lost when the summary is formed and how the analysis should be redesigned around that information. The canonical example is the generalized edge-count test, which keeps the two within-sample edge counts whose opposing deviations can cancel in the classical between-sample count. This paper traces how the same design question has led to task-specific methods built from graphs, graph-based ranks, kernels, and dissimilarity profiles for testing, change-point detection, covariate-balance assessment, clustering, classification, and generative-model evaluation.
\end{abstract}


{\small\noindent\textit{MSC2020 subject classifications.} Primary 62G10; secondary 62H15, 62H30.}

\section{Framing}

High dimensionality has long been associated with statistical difficulty---the familiar \emph{curse of dimensionality}, a phrase dating to Bellman (1957). A common response is to reduce dimension directly or to reduce effective complexity by imposing structure. Influential examples include principal component analysis (see Jolliffe, 2002), sufficient dimension reduction (Li, 1991), manifold learning (Roweis \& Saul, 2000; Tenenbaum, de Silva \& Langford, 2000), sparse regression (Tibshirani, 1996), and low-rank matrix recovery (Candès \& Recht, 2009). A separate literature has examined how high dimensions can help, often by exploiting concentration and separation phenomena (Donoho, 2000; Gorban \& Tyukin, 2018).

This paper presents a distinct perspective: the way relations among observations should be read can change with dimension. The use of relations in statistics is not new. In one dimension, the Wald--Wolfowitz runs test pools and orders the observations, then counts the runs in the resulting label sequence (Wald \& Wolfowitz, 1940). Its multivariate generalization by Friedman \& Rafsky (1979) replaces the ordering with a minimum spanning tree (MST) on the pooled observations. Schilling (1986) and Henze (1988) subsequently used $k$-nearest-neighbor graphs. Kernel two-sample tests instead aggregate pairwise similarities (Gretton et al., 2012). As dimension grows, however, new relational patterns can emerge, carrying useful signal that a conventional summary may miss. Cursive asks which of these emerging patterns matter for the task and how the statistical readout should change to use that information.

The name \emph{Cursive} is new; its underlying perspective emerged with the generalized edge-count test, posted as a preprint in 2013 and later published as Chen and Friedman (2017). That work identified a high-dimensional within/between-group asymmetry in a similarity graph and redesigned the statistic to prevent oppositely directed deviations in the two within-group edge counts from cancelling. The same design idea was later extended across representations and tasks, taking a form appropriate to the relations being used and the inferential target.

Cursive is therefore a method-generating research program rather than a single formula. Section~2 begins with the canonical generalized edge-count mechanism and then documents these extensions. Section~3 formulates the Cursive principle. Section~4 briefly contrasts these informative patterns with a harmful high-dimensional artifact before turning to the broader implications.

\section{Informative geometry: the curse reveals overlooked signal}

\subsection{The mechanism: a within/between-group asymmetry}

A clear statement of the mechanism is the generalized edge-count test (GET) of Chen \& Friedman (2017). For two samples $X$ and $Y$, the test constructs a similarity graph $G$ on the pooled observations (for example, a $k$-MST under Euclidean distance). Each edge belongs to one of three categories: \emph{within-$X$}, \emph{within-$Y$}, or \emph{between-sample}. The corresponding counts are denoted by
\[R_1 = \#\{\text{within-}X\text{ edges}\}, \quad R_2 = \#\{\text{within-}Y\text{ edges}\}, \quad R_0 = |G| - R_1 - R_2 \ (\text{between}).\]

The classical edge-count test (Friedman \& Rafsky, 1979) is based on a mixing argument. If the two samples come from the same distribution, their observations should be well mixed on the similarity graph. Under an alternative, observations are expected to lie closer to others from the same sample, so an unusually small number $R_0$ of between-sample edges provides evidence against the null.

As dimension grows, however, location and scale alternatives can produce qualitatively different patterns on the similarity graph. Under a location alternative, separation continues to make within-sample edges more prevalent and $R_0$ unusually small. Under a scale alternative, concentration of norms and pairwise distances sharpens a radial separation: the lower-variance sample forms an inner layer and the higher-variance sample an outer layer. A finite outer sample provides only sparse angular coverage of the high-dimensional outer shell, so many outer-layer observations find inner-layer observations closer than their available outer-layer neighbors and connect inward, while inner-layer observations tend to connect among themselves. The inner sample's within-count consequently tends to rise above its null expectation while the outer sample's within-count tends to fall below. $R_0$ remains near its null expectation when these opposing deviations largely cancel in $R_1+R_2=|G|-R_0$; if the outer sample's within-count decrease dominates, $R_0$ can instead rise above its null expectation. The net effect on $R_0$ depends on the two sample sizes, as well as on the scale contrast and the graph.

Chen \& Friedman (2017) quantify how densely the outer layer would have to be populated to overcome this geometry through a sphere-packing calculation: the number of well-separated points that can be placed on the surface of a $d$-dimensional unit ball grows astronomically with $d$ (on the order of $10^{10}$ at $d=30$, $10^{20}$ at $d=65$), sample sizes that cannot be reached in practice. Under scale alternatives, this geometry can therefore produce opposing deviations in $R_1$ and $R_2$ that are largely invisible to a statistic based only on the between-sample count $R_0$.

Their fix leaves the graph unchanged and changes how its edge pattern enters the test. The generalized statistic
\begingroup
\setlength{\abovedisplayskip}{4pt}
\setlength{\abovedisplayshortskip}{4pt}
\setlength{\belowdisplayskip}{5pt}
\setlength{\belowdisplayshortskip}{5pt}
\[S = (R_1 - \mu_1,\ R_2 - \mu_2)\,\Sigma^{-1}\,(R_1 - \mu_1,\ R_2 - \mu_2)^{\!\top},\]
\endgroup
with $\mu_i, \Sigma$ computed under the permutation null, measures how far $(R_1,R_2)$ departs from its joint null expectation, without restricting the direction of departure. It is therefore sensitive to location and scale alternatives, as well as to broader distributional differences. In particular, opposing within-group deviations under scale alternatives no longer cancel in the readout. Calibration is by the permutation null, which is finite-sample valid under exchangeability; an analytic $\chi^2_2$ approximation to that null performs well for sample sizes in the hundreds and makes the test computationally convenient for larger datasets. Figure~1 summarizes this mechanism for representative location and scale alternatives, showing how the classical and generalized statistics use the two within-sample deviations differently.

\begin{figure}[t]\centering
  \includegraphics[width=\linewidth]{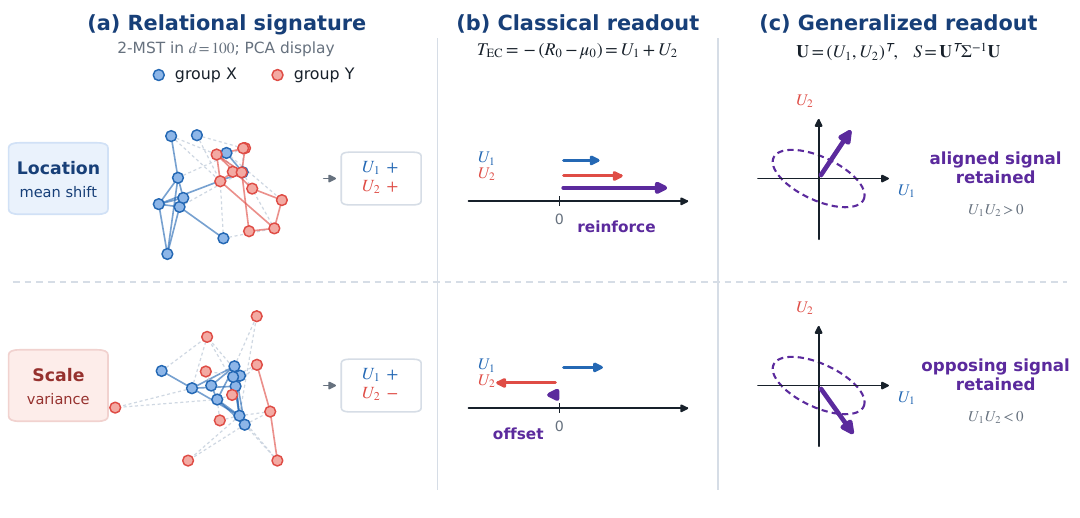}
  \caption{\textbf{The same relational representation, read in two ways.} Panel (a) shows location and scale Gaussian examples with $n_X=n_Y=10$ and $d=100$. In both, $X\sim N(0,I_d)$; $Y\sim N(0.3\mathbf 1,I_d)$ for location and $Y\sim N(0,1.2^2I_d)$ for scale. In each row, the pooled Euclidean 2-MST is constructed in the full $d$-dimensional space; node positions are a two-dimensional PCA projection used only for display. Solid coloured edges are within-group and dashed grey edges are between-group. Writing $U_i=R_i-\mathbb{E}_0R_i$, panel (b) shows that the classical readout $U_1+U_2$ reinforces the same-direction deviations under the location alternative but lets opposing deviations offset under scale alternatives. Panel (c) shows that $S=\mathbf U^\top\Sigma^{-1}\mathbf U$ retains both directions; the dashed ellipses are schematic level sets of $S$.}
\end{figure}

\noindent\begin{minipage}{\textwidth}
\setlength{\parindent}{1.5em}\indent
Figure~2 illustrates this contrast under the shrinking scale alternative $c=1+1/\sqrt d$, for which the marginal scale ratio approaches one as $d$ grows. The estimated power of the generalized statistic $S$ remains essentially one through $d=30000$, whereas the powers of $R_0$, the energy-distance test (Sz\'ekely \& Rizzo, 2013), and maximum mean discrepancy (MMD; Gretton et al., 2012) decline substantially as dimension grows.
\end{minipage}

\vfill
\noindent\begin{minipage}{\textwidth}
\centering
  \includegraphics[width=0.60\linewidth]{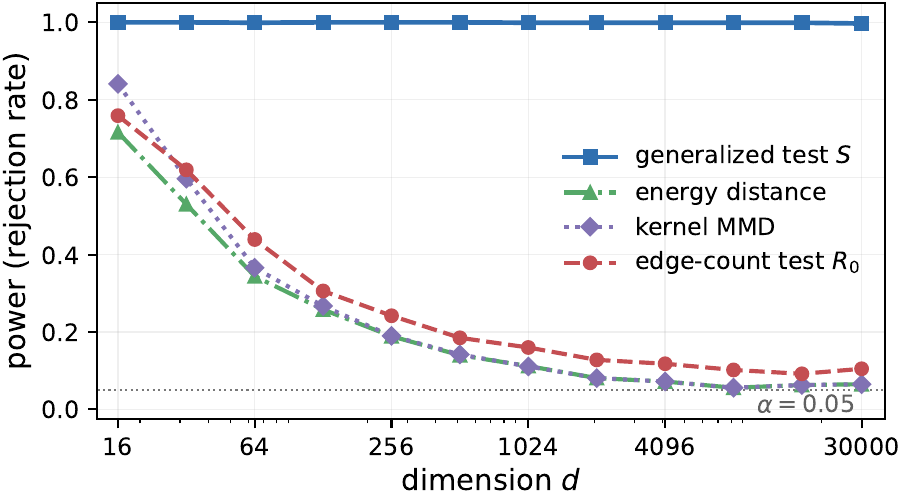}
  \captionsetup{hypcap=false}
  \captionof{figure}{\textbf{Power against dimension under a scale alternative.} Samples of size 100 are drawn from $N(0,\Omega)$ and $N(0,(1+1/\sqrt d)^2\Omega)$, where $\Omega_{ij}=0.5^{|i-j|}$. Rejection rates are based on 1,000 data sets at $\alpha=0.05$. The statistics $S$ and $R_0$ use a Euclidean 5-MST. All four tests use the same set of 199 label permutations within each data set; MMD uses the pooled median-distance bandwidth.}
\end{minipage}
\FloatBarrier

\subsection{Finding signal across representations}

In GET, the statistical change is in the readout; the graph still determines which relations are available to be read. Friedman \& Rafsky (1979) used the minimum spanning tree; Chen \& Friedman (2017) use a denser $k$-MST. The same readout can also be defined on other similarity graphs, such as a $k$-nearest-neighbor graph. Graph construction and statistical readout are distinct design choices: the former determines which relations enter the representation, whereas the latter determines how those relations are summarized.

The phenomenon is not tied to graphs: it also appears in several two-sample procedures based on distances, graph-induced ranks, and kernels. In the distance-based statistic of Biswas \& Ghosh (2014), the two sample-specific contrasts are squared before being added, so contrasts with opposite signs do not cancel.

Zhou \& Chen (2023) introduce two new graph-based ranks, each of which yields a rank-based two-sample readout. Constructed from a sequence of similarity graphs, a graph-induced rank records how persistently an edge appears across the sequence, whereas an overall rank orders the edge weights in the final graph. For either construction, the two within-sample rank sums enter the test jointly, so aligned and opposing movements in ranked similarity remain available rather than being combined in advance.

Kernels provide a parallel formulation: once the pooled observations are fixed, the cross-sample kernel average is determined by the two within-sample kernel averages, $\alpha$ and $\beta$. MMD reads a sample-size-weighted direction of this pair; for equal sample sizes it is their sum direction, paralleling the edge-count test's $R_1+R_2$ direction (Figure 1). Under scale alternatives, $\alpha$ and $\beta$ can move in opposing directions as dimension grows, causing their contributions to MMD to offset. The generalized permutation-based kernel test of Song \& Chen (2024a) instead combines their centered deviations through a covariance-standardized quadratic statistic, retaining departures in either direction.

The examples above vary the relational representation while keeping the two-sample task fixed. With a different task, the relevant relational components, their calibration, and their role in the decision may also change.

\subsection{Finding signal across tasks}

{\widowpenalty=10000
Change-point detection is the closest transfer across tasks. At each candidate split, the observations before and after the split are treated as two samples, and the generalized edge-count statistic is computed from the two within-segment edge counts. Scanning this statistic over candidate splits yields an offline procedure that is sensitive to both location and scale changes in high-dimensional data and remains applicable to non-Euclidean observations. The scan is asymptotically distribution-free and admits analytic $p$-value approximations (Chu \& Chen, 2019).

Rank and kernel constructions extend the same scan in different ways. RING-CPD replaces unweighted graph edges with graph-induced ranks and scans the resulting pair of within-segment rank sums, retaining aligned and opposing departures (Zhou \& Chen, 2025). GKCP instead scans the pair of within-segment kernel similarities, allowing both aligned and opposing movements to contribute; the latter can offset in an MMD scan (Song \& Chen, 2024b). KAP-CPD extends this construction across kernels: it keeps the two within-segment similarities separate for both a Gaussian kernel and a graphlet kernel, then combines the resulting four quantities through a joint covariance-standardized readout. The change-point decision can therefore use opposing movements within either kernel as well as complementary signals across the two kernels (Sun \& Chen, 2026).\par}

A separate line extends graph-based change-point detection from retrospective scanning to sequential monitoring. The initial online framework used the conventional nearest-neighbor edge count (Chen, 2019). Chu \& Chen (2022) showed that this statistic can lose power against scale changes as dimension grows and introduced generalized and max-type stopping rules, together with analytic average-run-length approximations.

The $K$-sample problem requires a larger set of relational components. Zhang et al. (2020) replace the pair $(R_1,R_2)$ by the vector of all $K$ within-group edge counts and form a $K$-dimensional quadratic statistic under the permutation null; group-specific deviations that could offset in the total within-group count are thereby retained. This test was used for pathway-level comparisons along cancer progression. For $K>2$, however, the within-group counts no longer determine the full between-group structure. Song \& Chen (2022b) therefore retain both the vector of $K$ within-group counts and the vector of $K(K-1)/2$ pair-specific between-group counts, or an equivalent linearly independent joint vector. The readout is enlarged to match the richer $K$-group geometry rather than inherited unchanged from the two-sample problem.

Covariate-balance assessment gives the two within-group connectivities a different inferential meaning. Good matching can make both counts unusually small; poor balance can make both large, while a scale imbalance can drive them in opposite directions as dimension grows. GET's quadratic readout treats all three patterns as departures from the permutation null and may therefore reject even when the two low counts indicate exceptional balance. CrossNN and CrossMST (Chen \& Small, 2022) instead use one-sided max-type readouts that reject when at least one within-group connectivity is unusually large, but not when both are unusually low.

Paired comparison retains the two within-sample edge counts, but pairing changes which graph edges can contribute and how the null distribution is generated. Under the paired null, the two observations within each pair are exchangeable, $(X_i,Y_i)\overset{d}{=}(Y_i,X_i)$, and calibration swaps them independently across pairs. An edge joining the two observations in the same pair therefore remains between-sample after every swap and cannot contribute to either within-sample count; only edges linking different pairs enter $R_1$ and $R_2$. Zhang, Chen \& Zhou (2027) derive the joint covariance of these counts under the resulting within-pair permutation distribution and use it to calibrate a quadratic readout. As in GET, variance differences can make the two counts move in opposite directions as dimension grows.

Independence testing compares relations among the $X$ observations with the corresponding relations among the $Y$ observations. Friedman \& Rafsky (1983) introduced a graph-theoretic test based on the similarity--similarity alignment. The generalized independence test of Liu, Zhou \& Chen (2024) broadens this relational readout by constructing similarity and dissimilarity relations separately for $X$ and $Y$, then forming four generalized correlations---dissimilarity--dissimilarity, dissimilarity--similarity, similarity--dissimilarity, and similarity--similarity---and reading them through a covariance-standardized quadratic statistic. The two mixed cases capture reverse dependence, in which observations similar in one variable tend to be dissimilar in the other; a similarity--similarity readout does not use this pattern.

{\clubpenalty=10000
In clustering, when groups differ in scale, directed nearest-neighbor relations can become asymmetric as dimension grows: observations in the higher-variance group often point to the lower-variance group, whereas the reverse occurs less often. Nearest-neighbor asymmetry clustering (NAC; Chen \& Lin, 2023) evaluates a candidate partition through two permutation-standardized statistics: a weighted sum of the two within-cluster counts records when both counts are large, while their difference records an imbalance between them. Its objective can use either statistic, allowing both location-separated and scale-separated cluster structure to support a partition.

For labelled data, Mo \& Chen (2023) use rank-transformed dissimilarity profiles (RTDP) to map an observation to a vector of class-wise mean ranks derived from pairwise dissimilarities, then classify the vector by QDA. This profile preserves how an observation relates to all classes rather than reducing those relations to a single closeness score. Location, scale, and broader distributional differences can create different orderings of these class-wise averages, including scale settings in which observations from the more-dispersed class lie closer to the concentrated class than to their own.

Finally, community detection provides a boundary extension of the Cursive design logic beyond dimension-driven interpoint geometry. A conventional modularity direction (Newman \& Girvan, 2004) handles assortative structure and, with its direction reversed, disassortative structure, but misses the opposing within-block deviations of a core--periphery pattern. Unified bigroups standardized edge-count analysis (UBSea; Lin \& Chen, 2023) forms standardized weighted-sum and difference statistics from the two within-block counts and uses them to select among the resulting mixing types. The same cancellation logic thus extends to directly observed network structure.\par}

\begin{table}[!htbp]
\centering\small
\caption{\textbf{Cursive across statistical tasks.} For each task, the table lists the relational construction, statistical readout, and corresponding references. Community detection is discussed separately as an extension to directly observed network structure and is therefore not included in the table.}
\label{tab:task-map}
\begin{tabularx}{\textwidth}{@{}>{\raggedright\arraybackslash}p{0.14\textwidth}>{\raggedright\arraybackslash}p{0.20\textwidth}>{\raggedright\arraybackslash}X>{\raggedright\arraybackslash}p{0.23\textwidth}@{}}
\toprule
Task & Relational construction & Statistical readout & Citation \\
\midrule
Two-sample & similarity graph; graph-induced rank; or kernel similarities & joint readout of the two within-sample quantities preserves aligned and opposing departures that can offset in the conventional between-edge or MMD direction & Chen \& Friedman (2017);\newline Zhang \& Chen (2022);\newline Zhou \& Chen (2023);\newline Song \& Chen (2024a);\newline Song \& Chen (2026) \\
\midrule
Change-point & graph, rank, or one or more kernels on a sequence & the two within-segment quantities are scanned over candidate splits or monitored sequentially; kernel aggregation retains the pair for each kernel before combination & Chu \& Chen (2019, 2022);\newline Song \& Chen (2022a);\newline Zhang \& Chen (2021);\newline Liu \& Chen (2022);\newline Song \& Chen (2024b);\newline Zhou \& Chen (2025);\newline Sun \& Chen (2026) \\
\midrule
$K$-sample & similarity graph on all $K$ samples & within-group and pair-specific between-group edge vectors preserve group- and pair-specific departures & Zhang et al. (2020);\newline Song \& Chen (2022b) \\
\midrule
Paired comparison & similarity graph on the paired data, with only edges between different pairs contributing & joint readout of two within-sample counts calibrated by within-pair swaps preserves opposing departures under variance differences & Zhang, Chen \& Zhou (2027) \\
\midrule
Covariate balance & nearest-neighbor graph or MST on treated and matched-control observations & one-sided criterion rejects unusually high within-group connectivity in either group; jointly low within-group connectivities, as produced by strong balance, do not trigger rejection & Chen \& Small (2022) \\
\midrule
Clustering & directed $k$-nearest-neighbor graph & weighted-sum statistic records when both within-cluster counts are large; difference statistic records an imbalance between them & Chen \& Lin (2023) \\
\midrule
Classification & class-indexed mean-rank profiles of pairwise dissimilarities & QDA uses the full class-wise profile, preserving ordering patterns created by location, scale, and broader distributional differences & Mo \& Chen (2023) \\
\midrule
Independence & similarity and dissimilarity relations in $X$ and $Y$ & covariance-standardized quadratic readout of all four generalized correlations, including the two mixed similarity--dissimilarity terms & Liu, Zhou \& Chen (2024) \\
\bottomrule
\end{tabularx}
\end{table}
\FloatBarrier

\subsection{A calibrated omnibus for generative-model evaluation}

Generative-model evaluation provides a worked example of combining several Cursive readouts to address a ranking problem. Two widely used metrics, Fr\'echet Inception Distance (FID; Heusel et al., 2017) and Kernel Inception Distance (KID; Bi\'nkowski et al., 2018), can produce misleading rankings. Chen (2026) shows that visually unrecognizable images matching a reference embedding's first two moments can receive a better FID than genuine held-out images; across controlled severity sweeps, FID can be flat or reversed, while KID has only weak rank association with severity for dispersion collapse and several moment-restored departure families.

The Z-resolved Integrated Diagnostic (ZID; Chen, 2026) addresses these ranking failures by forming six standardized components: two from the graph-rank two-sample test of Zhou \& Chen (2023) and four from the generalized kernel two-sample test of Song \& Chen (2024a), evaluated at two bandwidths. Within each pair, one component records aligned movements and the other opposing movements. Because different departures can concentrate signal in different components, the flat-Simes ranking score aggregates their magnitudes without diluting a strong response in one component by weaker responses in the others. Across controlled sweeps spanning location, dispersion, dependence, higher-order structure, multimodality, and off-manifold support, the score tracks increasing departure severity, including settings in which FID is flat or reversed; in the ImageNet matched-moment stress test, it assigns a smaller departure score to genuine held-out images and a much larger one to the visually unrecognizable optimized images, correcting the FID ordering.

\section{The Cursive principle}

The examples in Section~2 point to a common design principle. As dimension increases, new patterns can emerge in the relations among observations. The signal in such a pattern may be distributed across several components. In some cases, those components move in opposite directions, so combining them too early can obscure the signal. Cursive identifies the informative components and designs the task-specific procedure to preserve their distinct contributions.

The final output need not itself be vector-valued. The retained components may ultimately enter a quadratic statistic, a maximum, or another task-specific rule. GET, for example, combines its two within-group deviations through a scalar quadratic statistic. The relational construction determines which relations are available, whereas the statistical readout determines how their components enter the decision. The appropriate readout is therefore task-specific.

The choice of representation affects both the relations emphasized and the method's practical properties. Graphs can select a sparse set of local relations; rank constructions represent pairwise relations by their relative order; kernels quantify pairwise similarity according to the chosen kernel. These choices shape robustness, interpretability, and computation, and the most useful representation may differ across tasks.

\section{Discussion}

\subsection{A complementary problem: representation degradation}

The preceding sections have focused on using informative relational patterns that emerge as dimension increases. High dimension is better known, however, for the difficulties it creates. One such difficulty is degradation of the relational representation itself. In nearest-neighbor graphs, hubness can give a few observations disproportionate degree (Radovanović et al., 2010). Zhu \& Chen (2023) show that the resulting severe degree imbalance can erode the power of graph-based statistics and propose a robust-graph construction that balances proximity against a penalty on large vertex degrees. Paired with GET, the resulting reduction in degree variation can improve two-sample and change-point power, especially against scale differences, by reducing the permutation variance of $R_1-R_2$. This is an active repair of representation geometry degraded by the curse.

More generally, a constructed representation is not statistically neutral: the relational structure it exposes is shaped by both the underlying distribution and the construction itself. When the construction obscures task-relevant relations, redesigning the representation can enable a more effective readout.

\subsection{Implications}
\enlargethispage{\baselineskip}

The Cursive perspective is especially relevant when intrinsic dimension remains consequential for the task. Covering a unit $d$-dimensional region at resolution $\varepsilon$ requires on the order of $(1/\varepsilon)^d$ cells, already $10^{100}$ at $d=100$ and $\varepsilon=0.1$. Structure in the data can greatly reduce this worst-case burden, yet finite samples may still leave task-relevant regions sparsely covered. Even under such sparse coverage, informative patterns may remain visible in the relations among observations.

The work reviewed here shows how the insight behind GET extends across representations and tasks. As dimension grows, familiar statistical difficulties can arise, but so can informative relational patterns. The methods differ because the tasks differ, but they begin with the same question: which of these patterns carries useful signal for the task, and how should that signal shape the statistical decision? Designing statistics to read the trace left by the curse is Cursive.

\section*{Acknowledgments}

This work was supported in part by the National Science Foundation under Award No. DMS-2311399, \emph{Making Use of the Curse of Dimensionality in Modern Data Analysis}.

\section*{References}

\begin{itemize}[label={},leftmargin=1.4em,itemsep=2pt]
  \item Bellman, R. (1957). \emph{Dynamic Programming}. Princeton University Press.
  \item Bi\'nkowski, M., Sutherland, D. J., Arbel, M. \& Gretton, A. (2018). Demystifying MMD GANs. \emph{International Conference on Learning Representations}.
  \item Biswas, M. \& Ghosh, A. K. (2014). A nonparametric two-sample test applicable to high dimensional data. \emph{Journal of Multivariate Analysis} 123, 160--171.
  \item Candès, E. J. \& Recht, B. (2009). Exact matrix completion via convex optimization. \emph{Foundations of Computational Mathematics} 9, 717--772.
  \item Chen, H. (2019). Sequential change-point detection based on nearest neighbors. \emph{Annals of Statistics} 47(3), 1381--1407.
  \item Chen, H. (2026). What FID hides: detecting, ranking, and diagnosing deviations in generative evaluation. arXiv:2608.24881.
  \item Chen, H. \& Friedman, J. H. (2017). A new graph-based two-sample test for multivariate and object data. \emph{Journal of the American Statistical Association} 112(517), 397--409.
  \item Chen, H. \& Lin, X. (2023). High-dimensional clustering via nearest-neighbor asymmetry. arXiv:\allowbreak 2305.00578.
  \item Chen, H. \& Small, D. S. (2022). New multivariate tests for assessing covariate balance in matched observational studies. \emph{Biometrics} 78(1), 202--213.
  \item Chu, L. \& Chen, H. (2019). Asymptotic distribution-free change-point detection for multivariate and non-Euclidean data. \emph{Annals of Statistics} 47(1), 382--414.
  \item Chu, L. \& Chen, H. (2022). Sequential change-point detection for high-dimensional and non-Euclidean data. \emph{IEEE Transactions on Signal Processing} 70, 4498--4511.
  \item Donoho, D. L. (2000). High-dimensional data analysis: the curses and blessings of dimensionality. Invited lecture, Mathematical Challenges of the 21st Century, AMS National Meeting, Los Angeles.
  \item Friedman, J. H. \& Rafsky, L. C. (1979). Multivariate generalizations of the Wald--Wolfowitz and Smirnov two-sample tests. \emph{Annals of Statistics} 7(4), 697--717.
  \item Friedman, J. H. \& Rafsky, L. C. (1983). Graph-theoretic measures of multivariate association and prediction. \emph{Annals of Statistics} 11(2), 377--391.
  \item Gorban, A. N. \& Tyukin, I. Y. (2018). Blessing of dimensionality: mathematical foundations of the statistical physics of data. \emph{Philosophical Transactions of the Royal Society A} 376(2118), 20170237.
  \item Gretton, A., Borgwardt, K. M., Rasch, M. J., Schölkopf, B. \& Smola, A. (2012). A kernel two-sample test. \emph{Journal of Machine Learning Research} 13, 723--773.
  \item Henze, N. (1988). A multivariate two-sample test based on the number of nearest neighbor type coincidences. \emph{Annals of Statistics} 16(2), 772--783.
  \item Heusel, M., Ramsauer, H., Unterthiner, T., Nessler, B. \& Hochreiter, S. (2017). GANs trained by a two time-scale update rule converge to a local Nash equilibrium. \emph{Advances in Neural Information Processing Systems} 30, 6626--6637.
  \item Jolliffe, I. T. (2002). \emph{Principal Component Analysis}, 2nd ed. Springer.
  \item Li, K.-C. (1991). Sliced inverse regression for dimension reduction. \emph{Journal of the American Statistical Association} 86(414), 316--327.
  \item Lin, X. \& Chen, H. (2023). UBSea: a unified community detection framework. arXiv:2310.04934.
  \item Liu, M., Zhou, D. \& Chen, H. (2024). Generalized Independence Test. arXiv:\allowbreak 2409.07745.
  \item Liu, Y. \& Chen, H. (2022). A fast and efficient change-point detection framework based on approximate $k$-nearest neighbor graphs. \emph{IEEE Transactions on Signal Processing} 70, 1976--1986.
  \item Mo, X. \& Chen, H. (2023). Rank-transformed dissimilarity profiles for high-dimensional classification. arXiv:2306.15199.
  \item Newman, M. E. J. \& Girvan, M. (2004). Finding and evaluating community structure in networks. \emph{Physical Review E} 69, 026113.
  \item Radovanović, M., Nanopoulos, A. \& Ivanović, M. (2010). Hubs in space: popular nearest neighbors in high-dimensional data. \emph{Journal of Machine Learning Research} 11, 2487--2531.
  \item Roweis, S. T. \& Saul, L. K. (2000). Nonlinear dimensionality reduction by locally linear embedding. \emph{Science} 290(5500), 2323--2326.
  \item Schilling, M. F. (1986). Multivariate two-sample tests based on nearest neighbors. \emph{Journal of the American Statistical Association} 81(395), 799--806.
  \item Song, H. \& Chen, H. (2022a). Asymptotic distribution-free changepoint detection for data with repeated observations. \emph{Biometrika} 109(3), 783--798.
  \item Song, H. \& Chen, H. (2022b). New graph-based multi-sample tests for high-dimensional and non-Euclidean data. arXiv:\allowbreak 2205.13787.
  \item Song, H. \& Chen, H. (2024a). Generalized kernel two-sample tests. \emph{Biometrika} 111(3), 755--770.
  \item Song, H. \& Chen, H. (2024b). Practical and powerful kernel-based change-point detection. \emph{IEEE Transactions on Signal Processing} 72, 5174--5186.
  \item Song, H. \& Chen, H. (2026). A fast and effective kernel two-sample test for large-scale data. \emph{Statistics and Computing} 36(5), 214.
  \item Sun, M. \& Chen, H. (2026). KAP-CPD: Kernel aggregation for change-point detection in dynamic networks. arXiv:\allowbreak 2605.14463.
  \item Sz\'ekely, G. J. \& Rizzo, M. L. (2013). Energy statistics: a class of statistics based on distances. \emph{Journal of Statistical Planning and Inference} 143(8), 1249--1272.
  \item Tenenbaum, J. B., de Silva, V. \& Langford, J. C. (2000). A global geometric framework for nonlinear dimensionality reduction. \emph{Science} 290(5500), 2319--2323.
  \item Tibshirani, R. (1996). Regression shrinkage and selection via the lasso. \emph{Journal of the Royal Statistical Society: Series B} 58(1), 267--288.
  \item Wald, A. \& Wolfowitz, J. (1940). On a test whether two samples are from the same population. \emph{Annals of Mathematical Statistics} 11(2), 147--162.
  \item Zhang, J. \& Chen, H. (2022). Graph-based two-sample tests for data with repeated observations. \emph{Statistica Sinica} 32(1), 391--415.
  \item Zhang, J., Chen, H. \& Zhou, X.-H. (2027). A new non-parametric test for multivariate paired data from pair matching or paired designs. \emph{Journal of Statistical Planning and Inference} 246, 106443.
  \item Zhang, Q., Mahdi, G., Tinker, J. \& Chen, H. (2020). A graph-based multi-sample test for identifying pathways associated with cancer progression. \emph{Computational Biology and Chemistry} 87, 107285.
  \item Zhang, Y. \& Chen, H. (2021). Graph-based multiple change-point detection. arXiv:\allowbreak 2110.01170.
  \item Zhou, D. \& Chen, H. (2023). A new ranking scheme for modern data and its application to two-sample hypothesis testing. \emph{Proceedings of the 36th Conference on Learning Theory}, PMLR 195, 3615--3668.
  \item Zhou, D. \& Chen, H. (2025). Asymptotic distribution-free change-point detection for modern data based on a new ranking scheme. \emph{IEEE Transactions on Information Theory} 71(8), 6183--6197.
  \item Zhu, Y. \& Chen, H. (2023). Mitigating dimensionality effects with robust graph constructions for testing. arXiv:2307.15205.
\end{itemize}
\end{document}